# Diclofenac sodium ion exchange resin complex loaded melt cast films for sustained release ocular delivery

Goutham R. Adelli, Sai Prachetan Balguri, Prakash Bhagav, Vijayasankar Raman & Soumyajit Majumdar

RESEARCH ARTICLE

# Diclofenac sodium ion exchange resin complex loaded melt cast films for sustained release ocular delivery

Goutham R. Adelli[1], Sai Prachetan Balguri[1], Prakash Bhagav[1], Vijayasankar Raman[2], and Soumyajit Majumdar[1,3]

[1]Department of Pharmaceutics and Drug Delivery, School of Pharmacy, The University of Mississippi, University, MS, USA, [2]National Center for Natural Products Research, School of Pharmacy, The University of Mississippi, University, MS, USA, and [3]Research Institute of Pharmaceutical Sciences, The University of Mississippi, University, MS, USA

**Abstract**

*Purpose*: The goal of the present study is to develop polymeric matrix films loaded with a combination of free diclofenac sodium ($DFS_{free}$) and DFS:Ion exchange resin complexes (DFS:IR) for immediate and sustained release profiles, respectively.
*Methods*: Effect of ratio of DFS and IR on the DFS:IR complexation efficiency was studied using batch processing. DFS:IR complex, $DFS_{free}$, or a combination of $DFS_{free}$ + DFS:IR loaded matrix films were prepared by melt-cast technology. DFS content was 20% w/w in these matrix films. *In vitro* transcorneal permeability from the film formulations were compared against DFS solution, using a side-by-side diffusion apparatus, over a 6 h period. Ocular disposition of DFS from the solution, films and corresponding suspensions were evaluated in conscious New Zealand albino rabbits, 4 h and 8 h post-topical administration. All *in vivo* studies were carried out as per the University of Mississippi IACUC approved protocol.
*Results*: Complexation efficiency of DFS:IR was found to be 99% with a 1:1 ratio of DFS:IR. DFS release from DFS:IR suspension and the film were best-fit to a Higuchi model. *In vitro* transcorneal flux with the $DFS_{free}$ + DFS:$IR_{(1:1)}(1 + 1)$ was twice that of only DFS:$IR_{(1:1)}$ film. *In vivo*, DFS solution and DFS:$IR_{(1:1)}$ suspension formulations were not able to maintain therapeutic DFS levels in the aqueous humor (AH). Both $DFS_{free}$ and $DFS_{free}$ + DFS:$IR_{(1:1)}(3 + 1)$ loaded matrix films were able to achieve and maintain high DFS concentrations in the AH, but elimination of DFS from the ocular tissues was much faster with the $DFS_{free}$ formulation.
*Conclusion*: $DFS_{free}$ + DFS:IR combination loaded matrix films were able to deliver and maintain therapeutic DFS concentrations in the anterior ocular chamber for up to 8 h. Thus, free drug/IR complex loaded matrix films could be a potential topical ocular delivery platform for achieving immediate and sustained release characteristics.

## Definition

**Ion exchange resins** ↔ Water insoluble cross linked polymers with ionizable groups that can be exchanged to form complexes. Ion exchange resin (IR) in the present study represent Duolite™ AP 143/1083.
**DFS:$IR_{(1:1)}$** ↔ 1 part by weight of DFS is bound to 1 part by weight of IR.
**DFS:$IR_{(1:2)}$** ↔ 1 part by weight of DFS is bound to 2 part by weight of IR.
**DFS:$IR_{(2:1)}$** ↔ 2 part by weight of DFS is bound to or used to form complexes with 1 part by weight of IR.

$DFS_{free}$ **Film** ↔ Matrix film with unbound/free from of DFS without any IR.
**$DFS_{free}$ + DFS:$IR_{(1:1)}(1 + 1)$** ↔ 1 part by weight of DFS is bound to 1 part by weight of IR (for sustained release) and remaining 1 part by weight of DFS is in unbound or free state ($DFS_{free}$ for immediate release). For example, in a film with 1.6 mg of total DFS, 0.8 mg DFS is as DFS-IR and remaining 0.8 mg of DFS is as $DFS_{free}$.
**$DFS_{free}$ + DFS:$IR_{(1:1)}(3 + 1)$** ↔ 1 part by weight of DFS is bound to 1 part by weight of IR (for sustained release) and remaining 3 parts by weight of DFS is in unbound or free state ($DFS_{free}$ for immediate release). For example, in a film with 1.6 mg of total DFS, 0.4 mg DFS is as DFS-IR and remaining 1.2 mg of DFS is as $DFS_{free}$.

## Introduction

Ion exchange resins (IR) are water insoluble cross-linked polymers with ionizable groups that can be used to form complexes (Guo et al., 2009). They are differentiated into

Author for correspondence: Soumyajit Majumdar, Department of Pharmaceutics and Drug Delivery, School of Pharmacy, 111 Faser Hall, The University of Mississippi, University, MS 38677, USA. Tel: (662)-915-3793. Fax: (662)-915-1177. Email: majumso@olemiss.edu

anionic and cationic exchange resins based on the charge on the exchangeable ionic group (Sriwongjanya & Bodmeier, 1998). The strong cation exchangers, such as Amberlite™ IRP69, contain sulfonic acid functional groups, while the weak cation exchange resins, such as Amberlite™ IRP64 and IRP88, contain carboxylic acid functional groups. Similarly, anion exchange resins are also divided into strong exchange resins, with quaternary ammonium groups attached to the matrix such as Duolite™ AP 143/1083 (Figure provided in supplemental data) and Amberlite® IRA-410, and weak anion exchangers such as Dowex® WGR-2 and Amberlite® IRA-67.

In the past, IRs were primarily used in the field of agriculture and for the purification of water (Mantell, 1951). Application of IRs as excipients in the field of medicine started when synthetic ion-exchange resins were used as taste masking and as stabilizing agents in oral dosage forms (Kankkunen et al., 2002; Bhise et al., 2008; Patra et al., 2010; The Dow Chemical Company). Moreover, drug-IR complexes show modified release profiles when compared to the release of free drug (Halder & Sa, 2006). Saunders and Srivatsava studied the complexation efficiencies and release kinetics of alkaloids from IRs and suggested that IRs can be used as suitable carriers for the development of sustained-release formulations (Chaudhry & Saunders, 1956). Sriwongjanya & Bodmeier (1998) evaluated the complexation of propranolol hydrochloride and diclofenac sodium (DFS; Figure provided in supplemental data) with Amberlite® IRP 69 and Duolite® ATP 143, respectively. These drug-IR complexes were loaded in hydroxypropyl methyl cellulose (HPMC) matrix tablets to slow their release and to achieve a sustained release profile. Release rates depended on the amount, resin particle size and the type of carrier. The authors observed that the pH of the dissolution medium (0.1 N HCl or pH 7.4 phosphate buffer), or presence of the counter ion, had little to no effect on the release rate from the strong cation exchanger. With the weak cation exchange resin, *in situ* complex formation and retardation was only observed in pH 7.4 buffer but not in 0.1 N HCl because of the non-ionizable carboxyl groups. The use of smaller resin particles eliminated the burst release observed with the larger resin particles.

Modifying release rates has been one of the major applications of ion exchange resins in the pharmaceutical industry. Nicorette® is a widely used product for smoking cessation. It contains nicotine sorbed onto ion-exchange resin in a gum base (Ove et al., 1975). The drug–resinate offers a slower release profile for absorption over a 30-min period, aided by the mechanical chewing activity and the slow elution from the resin particles. Another example of controlled release application is a liquid suspension product of dextromethorphan called Delsym®. Dextromethorphan is bound to the ion-exchange resin and coated with ethylcellulose (Amsel, 1980). The bioavailability of the product is equivalent to that of dextromethorphan solution. Similarly, ocular films have been studied before to establish a sustained release profile employing their mucoadhesion mechanisms, hydration or degree of swelling of the polymers (Lee et al., 1999; Sasaki et al., 2003; Adelli et al., 2015). The uses of films with ion-exchange resins in combination with free/unbound drug and drug:IR complex, however, has not been studied before.

The use of IRs in the field of ophthalmic formulations, also has received little attention. Currently, Betoptic S® suspension containing 0.25% betaxolol HCl, is the only ophthalmic product available that utilizes ion exchange resins (Betoptic S®). In the present study, we are using drug-IR complexes loaded into polymeric melt cast films to deliver DFS into the eye for prolonged periods. DFS belongs to the class of drugs known as non-steroidal anti-inflammatory drugs (NSAIDs) and is used to treat swelling (inflammation) of the eye after cataract surgery (Herbort et al., 2000). DFS is also used after corneal refractive surgery (Fawzi et al., 2005) to temporarily relieve pain and photophobia (sensitivity to light). One drop of the currently marketed DFS ophthalmic solution formulation is applied to the affected eye, four times daily, beginning 24 h after cataract surgery and continued throughout the first two weeks of the postoperative period (Bausch & Lomb).

Drug delivery to the eye has always been a challenging task due to various physiological barriers (Lee & Robinson, 1986; Gaudana et al., 2010). Topical eye drops such as solutions, gels and suspensions are the most accepted and conventional drug delivery systems for treating ocular diseases (Sharma et al., 2016). However, the ocular barriers present major challenges when it comes to treating chronic issues such as chronic inflammation, dry eye, uveitis, age related macular degeneration, glaucoma and diabetic retinopathy (Edelhauser et al., 2010; Adelli et al., 2013). These conditions require long-term therapy and frequent administration of eye drops. Conventional drops deliver only 5–10% of the applied dose into the anterior segment of the eye because of precorneal drainage and other ocular barriers (Geroski & Edelhauser, 2000; Gaudana et al., 2010). As a result, the frequency of administration of the eye drops is usually 4-6 times a day, as in the case of DFS ophthalmic solution, based on the severity of the pathological condition. For posterior segment diseases, eye drops are mostly inefficient, so far, and although intravitreal injections are very effective, they are associated with complications such as pain, infection, endopthalmitis and retinal detachment (Edelhauser et al., 2010).

Thus, there is an unmet need for novel and efficient delivery strategies to prolong the duration of action of the drug or drug candidates in the eye. Some of the delivery systems currently under investigation include surface functionalized nanoparticles (Kompella et al., 2013; Suk et al., 2015; Wang et al., 2016), liposomes (Swaminathan & Ehrhardt, 2012; Honda et al., 2013; Agarwal et al., 2016), hydrogels (Li et al., 2013; Liang et al., 2016) and matrix films (Maichuk Iu & Iuzhakov, 1994; Kaur & Kanwar, 2002; Adelli et al., 2015; Maulvi et al., 2016). Amongst all the novel formulation strategies being explored, melt-extruded matrix films are the easiest to prepare and the fabrication process is free of solvents or other additives that might cause unwanted reactions at the site of application.

Thus, the objective of the present study is to evaluate DFS:IR complex (Figure provided in supplemental data) loaded polymeric matrix films for once or twice a day application. Matrix films loaded with DFS unbound (DFS$_{free}$) or DFS:IR complex or a combination of DFS$_{free}$ and DFS:IR complex (in various ratios) were examined with respect to DFS release profiles and ocular disposition following topical application. DFS ophthalmic solution (marketed) and corresponding IR suspensions were used as controls.

## Methods

### Chemicals

PEO [PolyOx® WSR N-10 (PEO N-10), MW: 100 000 Da; PubChem CID: 5327147] and Duolite™ AP 143/1083 (Cholestyramine Resin USP; PubChem CID: 70695641) were kindly donated by Dow Chemical Company (Midland, MI). DFS (PubChem CID: 5018304) was purchased from Sigma Aldrich (St. Louis, MO). Pearlitol® 160 C was obtained from Roquette Pharma as gift sample. DFS 0.1% w/v ophthalmic solution (Akorn Pharmaceuticals, Lake Forest, IL) was purchased from the pharmacy. All other chemicals were purchased from Fisher Scientific (St. Louis, MO).

### Animal tissues

Whole eye globes of New Zealand albino rabbits were purchased from Pel-Freez Biologicals® (Rogers, AK), shipped overnight in Hanks Balanced Salt Solution over wet ice (Majumdar et al., 2010) and used on the same day of receipt.

### Animals

Male New Zealand albino rabbits (2.0-2.5 kg) procured from Harlan Laboratories® (Indianapolis, IN) were used in all the studies. All animal experiments conformed to the tenets of the Association for Research in Vision and Ophthalmology statement on the Use of Animals in Ophthalmic and Vision Research and followed the University of Mississippi Institutional Animal Care and Use committee approved protocols (UM Protocol # 14-022).

### Preparation of DFS-IR complex

Prior to complex formation, IRs were micronized and sieved through a 400 US mesh size (Resin size < 37 microns) to avoid any burst release from larger particles (Sriwongjanya & Bodmeier, 1998). Then the resins were washed thoroughly with deionized water and activated by washing with chloride and hydroxide forms and then rinsed with water again. The drug-resin complexes were formed by batch process (The Dow Chemical Company, 2013). A concentrated aqueous solution of DFS (20 mg/mL) was prepared. To this, an accurately weighed amount of resin, three different ratios of DFS:IR = 1:2, 1:1 or 2:1, was added and agitated for 24 h. DFS:IR complexes (DFS:IR$_{1:2}$, DFS:IR$_{1:1}$ and DFS:IR$_{1:2}$) thus formed were separated out by centrifugation, washed with deionized water to remove unbound drug and dried in a desiccator overnight.

To calculate the bound drug percentage, the supernatant drug solution from the batch was collected before and after the agitation process. The amount of DFS present in the solution before and after the process were analyzed using the HPLC-UV method. The complexed resins were washed several times to remove any unbound or surface adsorbed DFS.

Percentage drug bound was calculated by analyzing the amount of drug remaining in the supernatant liquid and then using Equation (1):

$$\text{Percentage of bound drug} = \frac{A_1 - A_2}{A_1} \times 100 \quad (1)$$

where $A_1$ = amount of DFS in initial aqueous solution (mg); $A_2$ = amount of DFS in supernatant solution after 24 h (mg).

To confirm the complexation of DFS with IR, Fourier transmission infrared (FTIR) spectra for IR, DFS and DFS:IR complex were obtained using a Cary 660 series FTIR (Agilent Technologies, Santa Clara, CA) and MIRacle™ Single Reflection ATR (PIKE Technologies, Madison, WI).

### Preparation of polymeric matrix film and IR suspension

The melt-cast technology was employed in the preparation of the polymeric matrix film following a previously published protocol (Adelli et al., 2015). All matrix films were prepared with a 20% w/w DFS load. Briefly, DFS$_{free}$ and/or DFS:IR complex was mixed with PEO N10 (geometric dilutions) to obtain a uniform physical mixture. A 13 mm die was placed over a brass plate and heated to 70 °C using a hot plate. The physical mixture (200 mg) was added to the center of the die and compressed. The mixture was further heated for 2–3 min. After cooling, 4 mm × 2 mm sections, each weighing approximately 8 mg and with a drug load of 1.6 mg, were cut out from the film.

IR suspension was prepared by first dissolving the free DFS part in water (pH 7.2). Then, an accurately weighed amount of DFS:IR was added to the solution. Mannitol 4.5% w/v (Pearlitol® 160 C) was added as tonicity adjusting agent. To this, 0.5% HPMC (4000 cps) was added as a suspending agent and kept under stirring until all the HPMC dissolved. The pH of the formulation was adjusted to 7.3 using 0.1 N hydrochloric acid and/or sodium hydroxide.

### Assay and content uniformity

To determine the assay and content uniformity, a 1:1 mixture of isotonic phosphate buffered saline at pH 7.4 and dimethyl sulfoxide was used (The Dow Chemical Company, 2013). Each 8 mg film segment was placed in 50 mL of medium, to allow extraction of DFS from the resin, and sonicated for 15 min. This cloudy suspension was kept under constant stirring for 2 h, to allow complete release of the complexed drug, and centrifuged at 13 000 rpm for 15 min. The supernatant was filtered through 0.2 μ filter and analyzed for free drug using the HPLC-UV method.

### Scanning electron microscopy

The films (PEO N10, DFS$_{free}$ and DFS$_{free}$ + DFS:IR$_{1:1}$ (3 + 1)) were mounted on aluminum stubs using glued carbon tabs and then sputter coated for 120 s with gold using a Hummer 6.2 sputter coater (Anatech USA, Union City, CA). During the process, the gas pressure was at about 100 mTorr and the current was 15 mA. The surface morphology of the prepared samples was examined and digital micrographs were prepared using a JSM-5600 Scanning Electron Microscope (JEOL USA Inc., Peabody, MS) at an accelerating voltage of 5 kV.

### *In vitro* release and corneal permeability studies

To study the release profile of DFS from the various formulations, Slide-A-Lyzer dialysis MINI devices (10 000 Da MWCO) were used. A 20 mL glass vial was filled with 18 mL of release media (IPBS pH 7.34) and a magnetic stirrer was added to maintain equilibrium. Films (20%w/w DFS loading; Dose: 1.0 mg) loaded with DFS-IR$_{1:1}$ or with DFS$_{free}$ + DFS:IR$_{1:1}$(1 + 1) were placed in each dialysis device. Hundred microliters of IPBS was added in the dialysis device to wet the film. Similarly, in three more sets (n = 3), 1 mL of 0.1%w/v DFS ophthalmic solution (marketed formulation), DFS:IR$_{(1:1)}$ suspension and DFS$_{free}$ + DFS:IR$_{1:1}$(1 + 1) suspension (Dose: 1 mg) was added, to delineate the effect of the polymer matrix on the release profile. The glass vials were placed over a magnetic plate. The temperature was maintained at 34 ± 2 °C using calibrated hot plates. Aliquots, 0.8 mL, were collected at specific time intervals and replaced with an equal volume of release media. Studies were carried out for a period of 24 h. Samples were analyzed using a HPLC-UV method. The results from the studies were fit into zero order, first order, Higuchi and Boyd model.

Similar studies were carried out with DFS$_{free}$ + DFS:IR$_{1:1}$(1 + 1) to evaluate the effect of temperature (34 and 25 °C) and pH (0.1 N HCl, and water) and compared with the release profiles of DFS in IPBS pH 7.34 at 34 ± 2 °C. These studies were carried out for 12 h.

*In vitro* corneal flux and permeability of DFS from the DFS:IR complex loaded film formulation was evaluated using a side-by-side diffusion apparatus (PermeGear, Inc., Hellertown, PA) over a period of 6 h. The studies were carried out by sandwiching the film (4 mm × 2 mm; 20%w/w DFS; weighing 8 mg approximately; Dose: 1.6 mg) in between a Spectra/Por® membrane (MWCO: 10 000 Da) and isolated rabbit cornea (Pel-Freez Biologicals; Rogers, AK). Corneas were excised from whole eye globes, following previously published protocols (Majumdar et al., 2009, 2010). The membrane-film-cornea sandwich was then positioned in between the side-by-side diffusion cells (the chamber towards the Spectra/Por® membrane representing the periocular surface and the chamber towards the cornea representing the aqueous humor). Three milliliters (3 mL) of phosphate buffer, pH 7.4, was added to the periocular side and 3.2 mL was added to the aqueous humor side. Both chambers were sampled to evaluate the periocular loss and corneal permeation.

A different set-up was used to study transcorneal flux from the suspension and solution formulations. In this case, the cornea was mounted in between the two half-cells (membrane was not used in this case), 3 mL of DFS control solution or DFS:IR$_{(1:1)}$ complex suspension was added to the donor chamber. Phosphate buffer was used as receiver media (3.2 mL).

In all cases, the side-by-side diffusion cells were maintained at 34 °C using a circulating water bath. Six hundred microliters aliquots were collected from the receiver chamber(s) at specific time intervals and analyzed using the HPLC method.

### *In vivo* bioavailability studies

Male New Zealand albino rabbits weighing between 2.0 and 2.5 kg were used to determine *in vivo* ocular bioavailability of DFS from the topically instilled formulations. In these studies, 20%w/w DFS$_{free}$ and/or DFS:IR complex loaded films (weight 8 mg; dose: 1.6 mg) were placed in the conjunctival sac of the rabbit eye, while 1.6%w/v DFS:IR suspension (volume: 0.1 mL and dose: 1.6 mg) was administered in the eye. Hundred microliters of the 0.1% w/v DFS ophthalmic solution was administered twice with half an hour gap between the two applications (at -30 and 0 min; total dose: 200 µg).

Initially, studies were carried out for a period of 4 h post-application, with 0.1% w/v DFS ophthalmic solution, DFS$_{free}$ + DFS:IR$_{(1:1)}$(3 + 1) film and DFS$_{free}$ film. DFS ophthalmic solution (0.1% w/v) was used to understand ocular bioavailability of DFS from the solution formulation. Based on the 4 h data, another set of *in vivo* studies, evaluating ocular tissue concentrations 8 h post-topical administration, was undertaken using the following formulations: DFS$_{free}$ + DFS:IR$_{(1:1)}$(3 + 1) suspension, DFS$_{free}$ + DFS:IR$_{(1:1)}$(3 + 1) matrix films and DFS$_{free}$ loaded films. The compositions of all the formulations used in the *in vivo* studies are presented in Table 1. At the end of 4 h and 8 h, post-application of the last dose, the rabbits were euthanized under deep anesthesia with an overdose of pentobarbital injected through the marginal ear vein. The eyes were washed with ice cold IPBS and immediately enucleated and washed again. Ocular tissues were carefully isolated, weighed and preserved at −80 °C until further analysis.

Table 1. Composition of various formulations of Diclofenac sodium (DFS) used for ocular disposition studies.

| Formulations | Formulation #1 | Formulation #2 | Formulation #3 | Formulation #4 |
| --- | --- | --- | --- | --- |
| Ingredients | 0.1% DFS ophthalmic solution | DFS$_{free}$+DFS:IR$_{(1:1)}$(3 + 1) suspension | DFS$_{free}$ film | DFS$_{free}$+DFS:IR$_{(1:1)}$(3 + 1) film |
| Diclofenac sodium | 10 mg | 120 mg | 1.6 mg | 1.2 mg |
| DFS-IR::1:1 complex | – | 80 mg | – | 0.8 mg |
| Boric acid | √ | – | – | – |
| Edetate Disodium | 10 mg | – | – | – |
| Polyxyl 35 Castor Oil | √ | – | – | – |
| Sorbic acid | 20 mg | – | – | – |
| Tromethamine | √ | – | – | – |
| Mannitol | – | 450 mg | – | – |
| HPMC | – | 10 mg | – | – |
| PEO N10 | – | – | 6.4 mg | 6 mg |
| Water | 10 mL | 10 mL | – | – |

## Analytical procedure for in vitro samples

Waters HPLC system with 600 E pump controller, 717 plus auto sampler and 2487 UV detector was used. Data handling was carried out using an Agilent 3395 integrator. A 40:60 mixture of water (pH 3.5–4.0) and ACN was used as the mobile phase with Phenomenex *Luna*® 5 μm $C_{18}$ 100 Å, 250 × 4.6 mm column at a flow rate of 1.5 mL/min and 276 nm. DFS stock solution was prepared in mobile phase.

## Bio-analytical method

### Standard solution preparation

To 100 μL of aqueous humor (AH) or 500 μL of vitreous humor (VH) and to a weighed amount of the cornea, sclera, iris ciliary bodies (IC) and retina-choroid (RC) tissues, 20 μL of DFS stock in mobile phase (0.5, 1, 2.5, 5, 7.5 and 10 μg/mL) was added, vortexed and allowed to stand for 5 min. To precipitate the proteins, ice cold ACN was added to the AH and VH standards in a 1:1 ratio while 1 mL ACN was added to the cornea, sclera, IC and RC standards. Final concentrations of the standard solutions were in the range of 10–200 ng/mL for AH; 10–100 ng/mL for VH; 20–200 ng/mL: cornea and sclera and 10–200 ng/mL: IC and RC. Blanks were prepared, to test for specificity, for all the tissues by adding 20 μL of mobile phase instead of standard stock solutions to the respective tissues and following the same protocol. All samples were centrifuged at 13 000 rpm and 4 °C for 30 min and the supernatant was analyzed using HPLC-UV. All the standard curves generated an $R^2$ value greater than 0.97.

### Sample preparation

Approximately, 0.1 mL of AH and 0.5 mL of VH was collected from each test eye into individual centrifugal tubes. All other tissues, RC, IC, cornea and sclera, from each test eye were collected and weighed. Tissues were cut into very small pieces and placed into individual vials. Sample preparation and protein precipitation were carried out similar to the standard solution preparation protocol and the supernatant was analyzed using the HPLC-UV method.

A mixture of water (pH 3.5) and ACN in a ratio of 65:35 was used as the mobile phase with Phenomenex *Luna*® 5 μm $C_{18}$ 100 Å, 250 × 4.6 mm column at a flow rate of 1 mL/min and 284 nm.

## Data analysis

All experiments were carried out at least in triplicate. DFS release data were fitted to zero order, first order, Higuchi models and Boyd models (Equations 2, 3, 4 and 5).

$$C_t = C_0 + K_0 t \quad (2)$$

$$\text{Log} C_t = \text{Log} C_0 + Kt/2.303 \quad (3)$$

$$C_t = K_H t^{1/2} \quad (4)$$

where,

$C_0$ and $C_t$ = concentration at time 0 min and $t$ min; $K_0$, $K$ and $K_H$ = kinetic constants for zero order, first order and Higuchi models.

$$-\ln(1-F) = 3Pt/r \quad (5)$$

where, $P$ is the apparent permeability of the film, $F$ is fraction of drug released after time $t$ and $r$ is the thickness of the resin particles. The plot $-\ln(1-F)$ versus $t$ provides a linear line with a kinetic constant and coefficient correlation ($R^2$).

Drug diffusion parameters across cornea such as rate ($R$) and flux ($J$) were calculated using previously described method (Majumdar & Srirangam, 2009).

Statistical analysis was carried out using ANOVA to compare between different groups and Tukey's *post-hoc* HSD was used to compare differences between two groups. A $p$ value less than 0.05 was considered to denote statistically significant difference.

## Results

### Complexation efficiency

DFS was complexed at different ratios with Duolite™ AP 143 (1:1, 1:2 and 2:1). When the amount of DFS:IR was 2:1, only 49.3 ± 3.9% of DFS complexation was attained. At 1:1 and 1:2 ratios of DFS:IR a complexation efficiency of 99.03 and 99.04%, respectively, was achieved in both cases. Thus, all further studies using complexed DFS employed a 1:1 ratio of DFS and IR (DFS:IR$_{(1:1)}$).

FTIR spectra shows DFS binds to Duolite™ AP143 (Figure 1). Characteristic peak of DFS at 3430.5 cm$^{-1}$ (N-H stretching), 1573.5 cm$^{-1}$ (N-H bending) and 748.12 cm$^{-1}$ (C-Cl stretching) were observed in the FTIR spectra. Duolite™ AP 143 displays broad peak at about 3400 cm$^{-1}$ corresponding to the quaternary ammonium bending vibration, peaks at 2850 to 2900 cm$^{-1}$ were corresponding to CH, CH$_2$ and CH$_3$ stretching vibrations and two bands at 1600 and 1500 cm$^{-1}$ were corresponding to the aromatic ring. The N-H bending and C-Cl stretching were observed in DFS-IR complex demonstrating no covalent interaction between the drug and IR. The characteristic N-H stretching of DFS and the quaternary ammonium bending of Duolite resin can be observed in the DFS:IR spectra indicating the complexation between the resin and the drug (Figure 1).

### Assay and content uniformity

DFS content in all the formulations was between 94 and 103% of the theoretical values. DFS was found to be uniformly distributed within the matrix film (RSD < 2.3%).

### Scanning electron microscopy

Scanning electron microscopy images of the PEO N10, DFS$_{free}$ and DFS$_{free}$ + DFS:IR$_{1:1}$(3 + 1) films did not show any significant differences. The pores observed in the films facilitated the entry of water or the dissolution media for easy and rapid disintegration of the films (Figure 2).

### *In vitro release and corneal permeability studies*

Release of DFS, across the membrane, from the 0.1% w/v ophthalmic solution was 80% within 6 h. Percent release of DFS from DFS:IR$_{(1:1)}$ complex and DFS$_{free}$ + DFS:IR$_{(1:1)}$ (1 + 1) suspensions was 48.8 ± 2.3 and 72.4 ± 2.9, respectively, in 24 h. With the DFS:IR$_{(1:1)}$ complex and DFS$_{free}$ + DFS:IR$_{(1:1)}$(1 + 1) loaded matrix films, 52.7 ± 4.9 and 75.5 ± 3.8 percent of the DFS, respectively, was released across the membrane in 24 h (Figure 3). DFS release kinetics from each formulation is presented in Table 2. Percentage release of DFS from $DFS_{free}+DFS:IR_{(1:1)}(1+1)$ loaded matrix films in 0.1 N HCl and water were $0.7\pm0.07\%$ and $1.09\pm0.7\%$, respectively. Percentage release of DFS from $DFS_{free}+DFS:IR_{(1:1)}(1+1)$ loaded films at 25 °C in IPBS was $23.0\pm2.2\%$ compared to $60.2\pm3.5\%$ release at 34 °C at the end of 12 h.

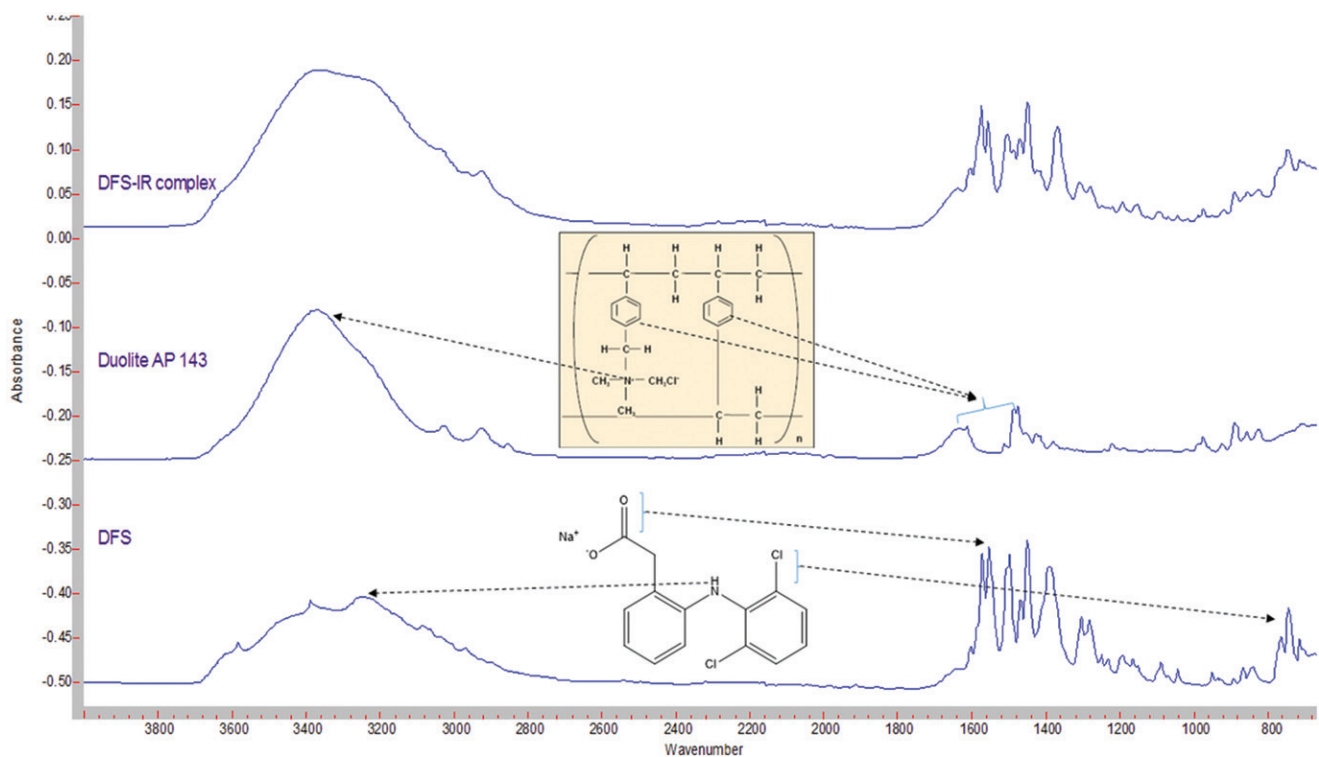

Figure 1. FTIR spectra of DFS, Duolite™ AP 143/1083 (IR) and DFS:IR complex.

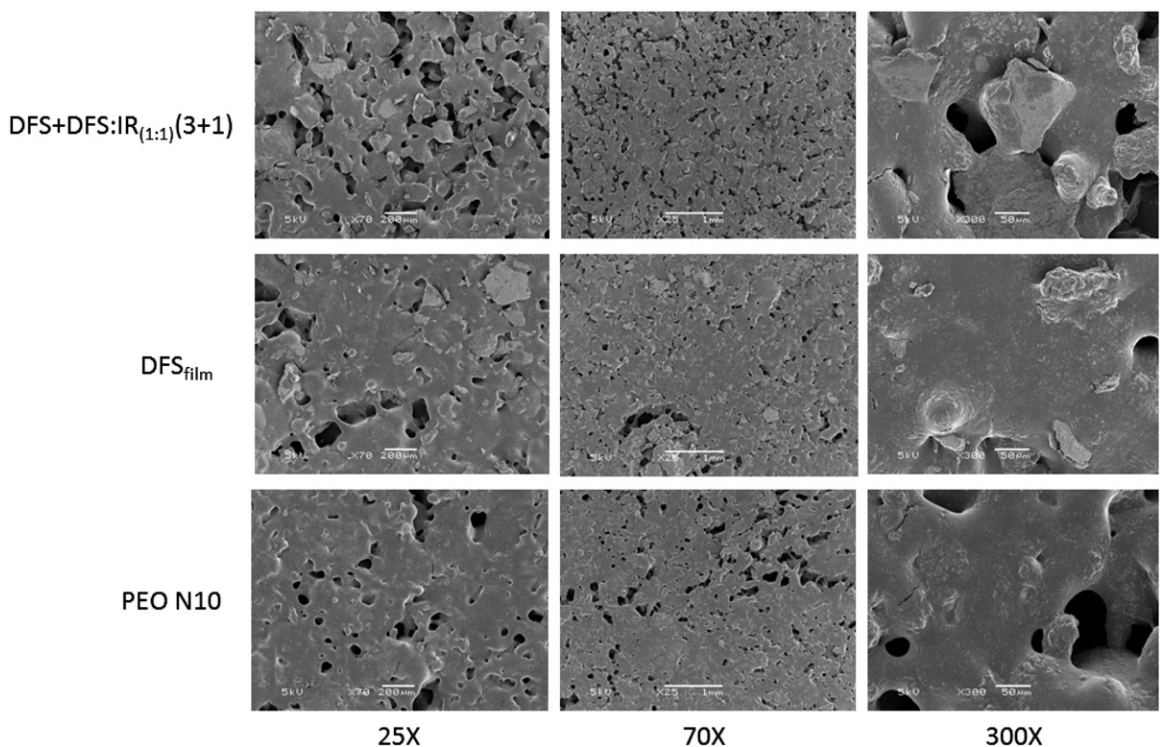

Figure 2. SEM of PEO N10, $DFS_{free}$ and $DFS_{free}+DFS:IR_{1:1}(3+1)$ films at 25×, 70× and 300× magnification.

Transcorneal flux from the various formulations was $10.2\pm0.2$ (0.1% w/v DFS ophthalmic solution), $2.0\pm0.9$ ($DFS_{free}$ matrix film), $0.34\pm0.04$ ($DFS:IR_{(1:1)}$ suspension), $0.4\pm0.05$ ($DFS:IR_{(1:1)}$ film) and $0.7\pm0.04$ ($DFS_{free}+$ $DFS:IR_{(1:1)}$ (1:1) film) μg/min/cm². Rate of permeability of DFS from the various formulations was $6.5\pm0.1$ (0.1% w/v DFS ophthalmic solution), $1.3\pm0.05$ ($DFS_{free}$ matrix film), $0.24\pm0.02$ ($DFS:IR_{(1:1)}$ suspension), $0.26\pm0.04$ ($DFS:IR_{(1:1)}$ film) and $0.46\pm0.02$ ($DFS_{free}+DFS:IR_{(1:1)}$ (1:1) film) μg/min.

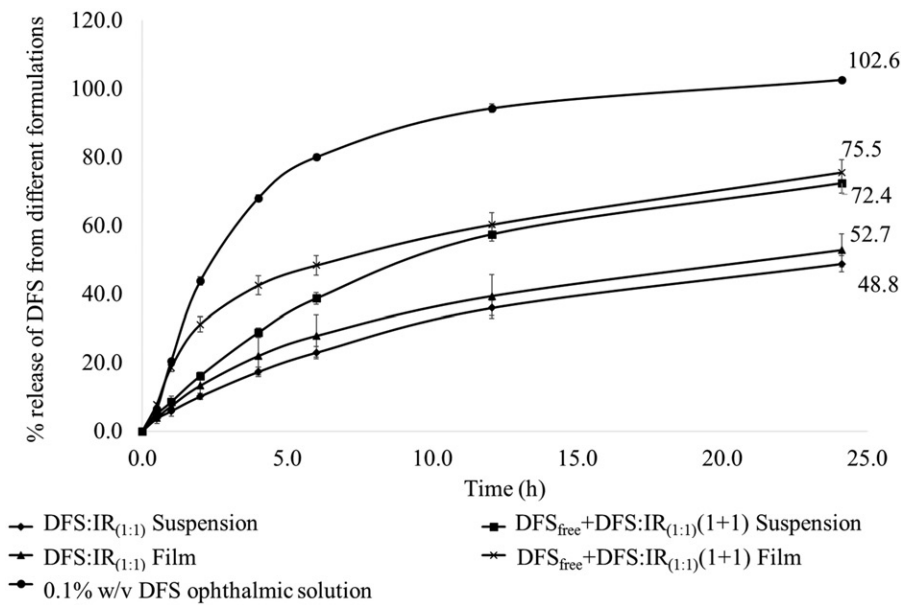

Figure 3. Percentage release of DFS up to 24 h form various ion exchange resin formulations and marketed ophthalmic solution.

Table 2. Model parameters obtained from fitting DFS release data to zero order, first order, Higuchi and Boyd models.

| Formulation | Rate order | Zero | First | Higuchi | Boyd |
| --- | --- | --- | --- | --- | --- |
| $DFS:IR_{(1:1)}$ Suspension | Kinetic constant | 2.0547 | 0.012 | 10.891 | 0.1195 |
|  | Coefficient correlation | 0.9273 | 0.9688 | 0.9894 | 0.9903 |
| $DFS_{free}+DFS:IR_{(1:1)}(1+1)$ Suspension | Kinetic constant | 3.0905 | 0.0237 | 16.681 | 0.1393 |
|  | Coefficient correlation | 0.8874 | 0.9671 | 0.9871 | 0.9959 |
| $DFS:IR_{(1:1)}$ Film | Kinetic constant | 2.1577 | 0.013 | 11.639 | 0.119 |
|  | Coefficient correlation | 0.8973 | 0.9573 | 0.9913 | 0.9949 |
| $DFS_{free}+DFS:IR_{(1:1)}(1+1)$ Film | Kinetic constant | 2.8066 | 0.0227 | 15.879 | 0.1294 |
|  | Coefficient correlation | 0.7871 | 0.9424 | 0.9567 | 0.9751 |
| 0.1% w/v ophthalmic solution | Kinetic constant | 3.9885 | 0.0961 | 23.449 | 0.1864 |
|  | Coefficient correlation | 0.6748 | 0.9895 | 0.8857 | 0.9687 |

DFS flux across the Spectra/Por® membrane (representing the precorneal loss) was $3.0\pm0.2$ and $5.5\pm0.6$ μg/min/cm² from the $DFS:IR_{(1:1)}$ film and $DFS_{free}+DFS:IR_{(1:1)}(1+1)$ film, respectively, under the settings employed.

### In vivo studies

Initially, ocular tissue concentrations were evaluated 4 h post-topical application of the DFS formulations. Ocular tissue DFS concentrations were analyzed only in the anterior segment of the eye (AH and IC) in this case. With the 0.1% w/v DFS ophthalmic solution, $181.4\pm64.8$ and $932.4\pm422.0$ ng of DFS/gm of tissue was detected in the AH and IC, respectively, at the end of 4 h. $DFS_{free}$ films was able to deliver $2.68\pm0.2$ μg and $2.7\pm0.06$ μg DFS/gm of tissue to the AH and IC, respectively. $DFS:IR_{(1:1)}$ complex loaded films delivered lower amounts of DFS to the anterior segment of the eye ($0.9\pm0.2$ and $1.02\pm0.05$ μg of DFS/gm of tissue to AH and IC, respectively) due to presence of DFS in the complexed form only.

To evaluate the sustained release effect of the matrix film with the DFS-IR complex, an 8 h study was undertaken. The ocular tissue concentrations obtained with the three formulations tested are presented in Table 3. At an equivalent dose, the IR suspension formulation delivered much lower concentrations than the IR-loaded film. DFS levels were below the limit of detection in the VH with all the formulations tested. Results from in vivo studies are presented in Table 3.

### Discussion

Use of IRs in the field of drug delivery as taste masking and controlled release excipients has been reported. Recently, IRs have also been used in ophthalmic dosage forms to achieve sustained release profiles and improved bioavailability. Jani et al. (1994) developed a suspension formulation of betaxalol hydrochloride (Betoptic® S) by binding it to IRs. Betoptic® S retards drug release in the tear, increases retention at the ocular surface and enhances drug bioavailability. As a result, Betoptic® S 0.25% is found to be bioequivalent to Betoptic Solution 0.5% in terms of lowering of intraocular pressure.

Table 3. Ocular tissue DFS concentration (μg/g of tissue) obtained from various topical formulations of DFS.

| Formulations (#) | Time (h) post instillation | Group # | Cornea | AH | IC | RC | Sclera |
| --- | --- | --- | --- | --- | --- | --- | --- |
| 0.1%w/v DFS solution (#1) | 4 | 1 | NA | 0.18 ± 0.06 | 0.93 ± 0.4 | NA | NA |
| 1.6%w/v DFS$_{free}$+DFS:IR$_{(1:1)}$(3 + 1) suspension (#2) | 8 | 2 | 1.3 ± 1.0 | 0.18 ± 0.07 | 0.43 ± 0.19 | ND | 8.4 ± 1.7 |
| 20%w/w DFS$_{free}$ Film (#3) | 4 | 3 | NA | 2.6 ± 0.2* | 2.7 ± 1.1$^b$ | NA | NA |
|  | 8 | 4 | 7.5 ± 3.4 | 0.7 ± 0.06$^b$ | 0.78 ± 0.35$^c$ | ND | 14.6 ± 4.8 |
| 20%w/w DFS$_{free}$+DFS:IR$_{(1:1)}$(3 + 1) Film (#4) | 4 | 5 | NA | 0.9 ± 0.2* | 1.02 ± 0.05 | NA | NA |
|  | 8 | 6 | 2.3 ± 1.4 | 0.36 ± 0.06$^a$ | 0.66 ± 0.15$^c$ | 0.09 ± 0.02 | 7.02 ± 4.03 |

AH - Aqueous humor, IC - Iris ciliary bodies, RC - Retina-choroid. NA - not analyzed, ND - below detection limit. Statistical significance was calculated using Tukey HSD (IBM SPSS 23). *Significantly different from all groups, $^a$Significantly different from group #5, $^b$Significantly different from all groups except group #5, $^c$Significantly different from group #3.

The goal of this project was to develop a matrix film loaded with IRs to allow immediate and sustained release profiles, thus providing both loading and maintenance doses, by using a combination of free and complexed drug in the matrix film. In the present study, an 8 mg matrix film with a 20% w/w drug load (1.6 mg of drug) was prepared. Of this 1.6 mg, 3 parts (1.2 mg) is maintained in the uncomplexed state (DFS$_{free}$) for immediate release and the other 0.4 mg (one part) is in the complexed state (DFS:IR$_{(1:1)}$) to provide the sustained release profile. The matrix film polymer by itself adds to the sustained release profile. The amounts of the free and bound DFS forms loaded in the films can be modified to achieve the required drug release profiles.

Duolite™ AP143 resin is an insoluble, strongly basic, anion exchange resin, supplied as a dry powder. It is suitable for use in pharmaceutical applications, both as an active ingredient (for adsorption of toxic chemicals) and as a carrier for acidic (anionic) drugs (Guo et al., 2009). Since DFS is negatively charged in the solution, DFS forms a complex with the positively charged Duolite™ AP143. In the present study, we evaluated the complexation efficiency of DFS with the resin at varying weight ratios of DFS and IR. No significant difference was observed between DFS:IR$_{(1:1)}$ and DFS:IR$_{(1:2)}$ in terms of complexation efficiency. Thus, all further studies were carried out using the DFS:IR$_{(1:1)}$ complex. A combination of DFS$_{free}$ and DFS:IR$_{(1:1)}$ was evaluated to provide immediate and sustained release components.

In vitro release of DFS from DFS-IR complex formulations showed an immediate release followed by a sustained release profile. Eighty percent release was observed with the 0.1% DFS ophthalmic solution within 6 h (102% in 24 h). With DFS$_{free}$+DFS:IR$_{(1:1)}$(1 + 1) combination in film, unbound DFS was released in the first 2 h. Due to the equilibrium between the bound and complexed form, sustained release of the remaining DFS was observed from the complex in the later part of the release profile. The initial release rate of DFS from DFS$_{free}$+DFS:IR$_{(1:1)}$(1 + 1) suspension formulation was slower compared to that from the DFS$_{free}$+DFS:IR$_{(1:1)}$(1 + 1) film. This can be attributed to the greater concentration gradient with the film than the suspension formulation. In the film, the 1.6 mg of DFS is concentrated within a 4 mm × 2 mm surface area, while in a suspension it is distributed over a greater surface area, thus reducing the concentration gradient. DFS:IR$_{(1:1)}$ film and suspension showed approximately 33% less release compared to the DFS$_{free}$+DFS:IR$_{(1:1)}$(1 + 1) film and suspension. This is because the total DFS exists in the complexed state in the DFS:IR$_{(1:1)}$; the release was thus more sustained. The DFS:IR$_{(1:1)}$ film showed slightly higher release rates than the suspension, similar to the DFS$_{free}$+DFS:IR$_{(1:1)}$. The data were fitted to zero order, first order and Higuchi kinetic models. The 0.1% ophthalmic solution exhibited first-order release kinetics with a coefficient of determination ($R^2$) of 0.9895. All other formulations (DFS:IR$_{(1:1)}$ suspension, DFS:IR$_{(1:1)}$ film, DFS$_{free}$+DFS:IR$_{(1:1)}$ suspension and DFS$_{free}$+DFS:IR$_{(1:1)}$ film) were best fitted to the Boyd model with $R^2$ of 0.9903, 0.9959, 0.9949 and 0.9751, respectively (Table 2). According to Boyd et al. (1947), the drug release from ion exchange resinate can be controlled by two kinds of diffusion processes, namely the diffusion of the drug across the thin film at the periphery termed as film diffusion and the diffusion of the drug in the matrix termed as particle diffusion. The rate-controlling step is either diffusion of drug across a thin liquid film at the periphery of the resin particle or diffusion of freed drug in a matrix (Jeong et al., 2007). From Table 2, we can see that in case of DFS:IR$_{(1:1)}$ film diffusion gives better linearity indicating that it is the rate limiting step. On the contrary, DFS$_{free}$+DFS:IR$_{(1:1)}$(1 + 1) particle diffusion was observed to be the rate limiting step due to the presence of free drug.

Percentage release of DFS from DFS$_{free}$+DFS:IR$_{(1:1)}$(1 + 1) was very negligible in water and in 0.1N HCl due to the absence of counter ions in the release media. Percentage release of DFS at room temperature (25 °C) was less compared to the release at 34 °C because of polymer chain relaxation at higher temperature. This results in faster gelation and drug release.

In vitro transcorneal permeability studies were performed over a period of 6 h. With the DFS:IR$_{(1:1)}$ loaded film (total 1.6 mg in complexed state) flux across the isolated cornea was approximately half of the flux obtained from the DFS$_{free}$+DFS:IR$_{(1:1)}$(1 + 1) loaded film. Presence of DFS in the free state increased flux of DFS across the cornea. Similarly, DFS flux from the DFS$_{free}$ containing film was higher than that of the DFS$_{free}$+DFS:IR$_{(1:1)}$(1 + 1) film formulation. With the DFS$_{free}$+DFS:IR$_{(1:1)}$ film, flux was 1/3rd of the flux obtained from the DFS$_{free}$ film. This was because the total dose (1.6 mg of DFS) was in unbound state. In the case of the ophthalmic solution, the total DFS is in solution resulting in high flux compared to all other formulations.

Based on the *in vitro* release and permeability data, it was apparent that the $DFS_{free} + DFS:IR_{(1:1)}(1+1)$ formulations were retaining DFS for longer period of time - 50% DFS release occurred in 12 h. *In vivo* studies were thus carried out with three parts of $DFS_{free}$ (1.2 mg) and one part (0.4 mg) as $DFS:IR_{(1:1)}$, $DFS_{free} + DFS:IR_{(1:1)}(3+1)$, to allow faster release of DFS from the combination matrix system. When $DFS_{free} + DFS:IR_{(1:1)}(3+1)$ suspension was administered a slight increase in the blinking rate was observed compared to the ophthalmic solution. The $DFS_{free} + DFS:IR_{(1:1)}(3+1)$ suspension formulation caused slight discomfort to the rabbit eye. The polymeric matrix film formulations, however, did not induce excessive tearing, redness or allergies as reported in our previous study (Adelli et al., 2015). Ocular discomfort was not observed with the $DFS_{free} + DFS:IR_{(1:1)}(3+1)$ film also. This is probably because the particles were embedded in the film and, thus, did not cause any discomfort to the eye.

DFS ophthalmic solution was tested *in vivo* in the rabbits to determine the DFS levels obtained in the ocular tissues with the currently marketed formulations, in the experimental model. Expectedly, with the 0.1% w/v ophthalmic solution, DFS concentrations in the AH and IC were much lower at the end of 4 h compared to other formulations tested. $DFS_{free}$ films, on the other hand, produced the highest concentrations in the AH and IC. Concentration of DFS in AH from 0.1% w/v ophthalmic solution was found similar to the data presented by Li et al. (2012): approximately 0.1 µg at 4 h post-topical application of 50 µL of 3.65 mg/mL of DFS solution. The matrix film transforms into a gel and releases the free drug much faster than that from the $DFS:IR_{(1:1)}$ or the $DFS_{free} + DFS:IR_{(1:1)}(3+1)$ loaded films. As a result, we see high DFS concentrations in the AH and IC from the $DFS_{free}$ films post topical application (4 h). In contrast, when the $DFS_{free} + DFS:IR_{(1:1)}(3+1)$ loaded matrix film transforms into a gel it starts releasing the free fraction of DFS, and, simultaneously, the bound DFS is also exchanged for the counter ions in the tear fluid releasing DFS from the $DFS:IR_{(1:1)}$ complex. IRs have the ability to form *in situ* complexes (Sriwongjanya & Bodmeier, 1998). As a result, some of the free DFS again starts forming a complex to maintain the binding equilibrium. Because of this constant change in the equilibrium the ocular tissue levels obtained from the $DFS_{free} + DFS:IR_{(1:1)}(3+1)$ film were lower compared to the $DFS_{free}$ film.

$DFS_{free}$ and $DFS_{free} + DFS:IR_{(1:1)}(3+1)$ films were evaluated for ocular tissue distribution of DFS 8 h post-topical administration. In addition to the films, $DFS_{free} + DFS:IR_{(1:1)}(3+1)$ suspension (same dose) was also tested to delineate the effect of the polymeric matrix film on the disposition of DFS in the ocular tissues. When $DFS_{free} + DFS:IR_{(1:1)}(3+1)$ suspension was administered, AH DFS concentrations achieved were similar but DFS concentrations were lower in the IC. This could be because of higher precorneal loss with the suspension dosage form. At the end of 8 h, the $DFS_{free}$ film produced higher DFS levels in the AH and IC compared to the $DFS_{free} + DFS:IR_{(1:1)}(3+1)$ suspension or film. Importantly, the rate of elimination of DFS from the ocular tissues between the 4 h and 8 h time period, was much faster with the $DFS_{free}$ film compared to the $DFS_{free} + DFS:IR_{(1:1)}(3+1)$ film: 0.02375 and 0.0623 µg/h, respectively, from the AH. Similarly, the rate of elimination of DFS from IC for $DFS_{free}$ film and $DFS_{free} + DFS:IR_{(1:1)}(3+1)$ film were 0.2347 and 0.0441 µg/h, respectively. Since there is a constant release of DFS from $DFS:IR_{(1:1)}$, DFS absorption phase was extended. As a result, rate of elimination from these tissues was lower compared to the $DFS_{free}$ film.

With the $DFS_{free} + DFS:IR_{(1:1)}(3+1)$ suspension, the DFS concentration gradient build up was low and, thus, most of the DFS was restricted to the cornea and sclera, the outermost ocular tissues. At the end of 8 h, posterior segment ocular tissues were also evaluated for DFS concentrations. None of the formulations were able to deliver detectable DFS levels to the VH. The $DFS_{free} + DFS:IR_{(1:1)}(3+1)$ film was able to deliver low but detectable DFS levels to the RC ($89.2 \pm 24.2$ ng/g of tissue).

According to Blanco et al. (1999), half-maximal inhibitory concentration ($IC_{50}$) values of DFS for inhibiting COX-1 and COX-2 enzymes is 0.611 µM (194.4 ng/mL) and 0.63 µM (200.4 ng/mL), respectively. DFS concentrations obtained from both solution and suspension formulations in the AH were below $IC_{50}$ levels. Both $DFS_{free}$ and $DFS_{free} + DFS:IR_{(1:1)}(3+1)$ loaded films were able to maintain significant DFS levels in the ocular tissues even at the end of 8 h post-topical application. Additionally, rate of elimination from the inner ocular tissues with the DFS:IR film formulation was significantly slower compared to the $DFS_{free}$ films. Thus, $DFS:IR_{(1:1)}$ loaded matrix films can maintain DFS levels for prolonged periods of time and may allow at least twice a day application. Moreover, the DFS:IR loaded dosage forms avoids the DFS concentration spikes in the ocular tissues noted with the other dosage forms including the $DFS_{free}$ loaded matrix films.

## Conclusions

This is the first report, to the best of our knowledge, investigating the effectiveness of an IR complex loaded matrix film as a topical ocular drug delivery platform. Ocular tissue DFS concentrations obtained from the matrix films generated high concentrations in the AH and IC bodies. Although, $DFS_{free}$ film was able to produce high DFS concentrations in the ocular tissues, DFS:IR film showed more controlled release across the tissues. Interestingly, only the DFS:IR film was able to deliver the drug to the posterior segment of the eye (RC). Thus, drug-IR complexes loaded into a matrix film can serve as a perfect platform for both immediate and sustained release systems. Modification of the IR film using different resins, different ratios of free to bound drug concentrations and different melt-extrudable polymer types can be used to achieve the desired drug release profiles.

## Declaration of interest

The authors report no declarations of interest.

This publication was supported by Grant 1R01EY022120-01A1 from the National Institutes of Health. The content is solely the responsibility of the authors and does not necessarily represent the official views of the National Institutes of Health.

**Supplementary materials available online**